# A Markovian Model for Assessing the Consistency of Vehicular Storage Systems

Mauro Femminella, Gianluca Reali

Dipartimento di Ingegneria Elettronica e dell'Informazione (DIEI), University of Perugia, Via G. Duranti 92, Perugia, Italy
gianluca.reali@diei.unipg.it, mauro.femminella@diei.unipg.it

Walter Colitti, Kris Steenhaut

Dept. of Electronics and Informatics (ETRO), Vrije Universiteit Brussel, B-1050 - Brussel - Belgium
wcolitti@etro.vub.ac.be, ksteenha@etro.vub.ac.be

*Abstract*— **In this paper, we evaluate the suitability of the vehicular devices for supporting data storage and distribution applications. Vehicular storage systems have recently emerged as means for making the information related to the vehicular environment available to vehicular users. Information could either be collected by sensors mounted onboard vehicles or coming from the surrounding environment. Modeling a vehicular data storage and distribution system requires the introduction of a number of parameters, such as the vehicle concentration and transmission range, the value of which may determine the suitability or unsuitability of the vehicular environment for establishing a communication infrastructure with the desired quality. In this paper we show how to solve this problem by resorting to a Markovian model and some results achieved through its application.**

## I. INTRODUCTION

Although the vehicular access to communication networks has a tradition of decades, the use of the vehicular environment for establishing a communication infrastructure, which stores information and makes it available on request, is in its early stage. For example, the huge research on the introduction of reliable communication services in the vehicular environment has produced a significant change over the time in the semantic of vehicular cloud services. Initially they were regarded as the mere remote access to a cloud infrastructure from vehicular devices. Now they mean the establishment of cloud computing platforms within a vehicular network using the processing and storing capabilities of vehicular devices [4].

Cloud computing is a generic name referring to a wide range of services based on the massive use of distributed computing resources and data storage capacity over the Internet. In this paper we will refer to *storage vehicular clouds*, i.e. a coordinated use of data storage resources physically located in mobile devices, transported by users in motion within vehicles.

The possibility of referring to such an environment as a potential communication infrastructure is due to the increasing proliferation of smart-phones, equipped with any kind of peripherals, such as cameras, GPS receivers, noise sensors, and many other ones. In addition, they are equipped with a significant amount of memory, in the order of tens of Gbytes. Thus, they can be regarded both as a information source and data repository. In addition, the car industry has produced vehicles equipped with smart sensors which monitor key parameters affecting the vehicle movement. The information of interest in such an environment may include vehicle traffic congestion, or local weather condition, such as nearby fog banks. In addition, the information of the local environment could be of interest, such as tourist information or available rescue services. The distribution of this, so-called, location-based information, is regarded as a very important task.

For example, the IDEA project [7], funded the Belgian IWT governmental agency, focuses on environmental stressors that have a very local character, such as ultrafine particulate matter and noise. Both fixed and vehicular sensors can collect such information and make it available on request through a client-server protocol, where the server can be located at the sensor site. A further example consists of the so-called BikeNet, a mobile sensing system for mapping the cyclist experience [6].

In this paper we propose a Markovian model for analyzing a vehicular environment in order to assess its suitability for implementing a vehicular data storage systems. The usage of such a system is intended for making the information types mentioned above available to vehicular users on request.

In Section II we define the problem and describe the mathematical model. This model will be used to analyze some case studies, and the relevant results are shown in Section III. Some final considerations are illustrated in Section IV.

## II. PROBLEM DESCRIPTION AND STATISTICAL MODEL

A functional description of a vehicular communication model requires the identification of a set of key building blocks, that provide a foundation for the services to be deployed. They include:

- mechanisms and protocols for service description and implementation;
- architectural framework, able to provide suitable insights for the service operation, including coexistence with legacy solutions, dynamic deployment, and hot re-configurability;
- performance metrics, in terms of QoE, scalability, total resource utilization, and signaling load.

Nevertheless, the analysis shown in this manuscript has a more general scope. Our aim is to assess the suitability of a vehicular environment for establishing a reliable data storage system which makes information available on-request. Fig. 1 shows the situation considered.

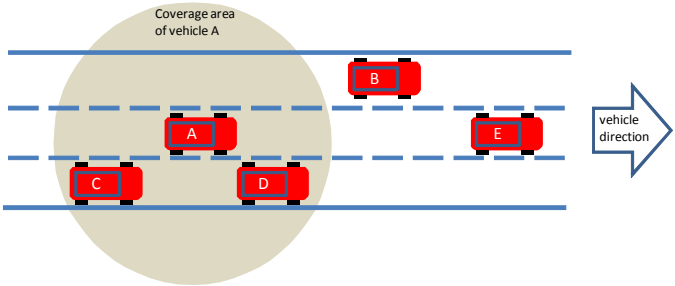

Fig. 1. Sketch of the vehicular environment considered.

A set of vehicles are moving along the indicated direction. We assume that vehicles are equipped with radio both transmission and reception capabilities implemented through any kind of available technologies, such as WiFi or WAVE [8]. We also assume that a piece of information, referred to as Information Record (IR), has to be made available to them within a given time time. A possible approach to achieve this objective is to install a number of fixed roadside units which can exchange data with vehicles. Although this approach is sustainable in local areas, it is hard to apply it in large areas. For example, to cover a highway for hundreds of kilometers by installing a roadside unit every 100 meters, along with the relevant network infrastructure, would be very expensive. An alternative solution is to install roadside units at a relative distance of some km, and resort to some vehicles for making information available to all other vehicles. A similar situation consists of a vehicular sensor which acquires information which is of potential interest for all vehicles moving in the same direction in the same area. For example if an oil spot on the road is detected, this IR should be communicated or, at least, made available to all vehicles located up to, for example, 1 km upstream.

The question is how to store any IR in the vehicular network so as to deliver it to any requesting vehicle which stays within the desired road segment.

In this paper we do not focus on the functional properties of the protocols suited for both inter-vehicle and road-to-vehicle communications. The relevant communication, synchronization, and signaling problems have been analyzed in many papers [2][3] and research projects [7]. What counts in our analysis is the ability of any vehicle to distribute its stored information to all other vehicles located within its coverage area. For example, in Fig. 1 it is assumed that the vehicle A has to distribute an IR. It can reach vehicles C and D. In turn, these vehicles can distribute that same information towards other vehicles. In this way the resulting information distribution takes place according to an *epidemic* distribution.

Our analysis focuses on the consistency of the vehicular environment, defined as follows:

*We define a vehicular network consistent, with respect to an information record to be made available within a given road segment of length L, if such a record is stored, at least, in a desired target number t of vehicles..*

The target number *t* is clearly application dependent. In case of an IR related to safety, *t* could be equal to almost *all* vehicles in the road segment considered. In case of an IR related to safety, t could be equal to almost all vehicles in the road segment considered. In case of an IR of less urgent nature, a number from 1 to some units or tens of vehicles, which make the IR available on request, could be satisfactory.

The final objective of our analysis is to determine the IR probability of consistency. We assume that the vehicles are randomly distributed over the road segment for which the IR is assumed to be meaningful. Such an IR is therefore not propagated outside the considered road segment. The knowledge of the specific mechanism used for keeping the IR within the desired road segment is not necessary for the forthcoming analysis. For this purpose many solutions are available, such as the inclusion of the GPS coordinates of the end points of the road segment within the IR.

The reader could observe that in case of high traffic, vehicles could form queues. In this case, it is evident that the high number of vehicles and their storage resources, which move slowly and close to each other, make the consistency probability close to 1 in a very short time. The challenging situation appears when the vehicles move quite freely along the considered stretch of road. In this case, both the number of vehicles in the road segment and their relevant distance are rapidly changing random variables and the desired probability of consistency within a given time frame is not easily achievable.

What is extremely important is the dynamic behavior of the data storage system. In particular, given some key parameters describing the situation under analysis, it is necessary to evaluate the time necessary for achieving a desired value of probability of consistency. This time (e.g. few seconds or tens of seconds) determines the class of services that can benefit from the vehicular data storage system. For this reason, the analysis in what follows focuses on both the transient and the steady state probability.

A statistical model, suitable for evaluating the resulting probability of consistency, must include all the events which determine the system's evolution. We will use the following symbols:

$L$: length of the road segment in which an IR must be made available.

$N$: number of vehicles present in the considered road segment.

$P$: communication range of each vehicle, corresponding to the diameter of its coverage area.

$N_{ave}$: average number of vehicles staying within a coverage area. If we assume that $P$ is much higher than the road width, we can compute $N_{ave}= PN/L$.

$v$: Average vehicle speed.

$T=L/v$: average vehicle sojourn time within the considered road segment. It follows that $1/T=\mu$ is the departure rate of vehicles. From the Little law it also follows that $N\cdot\mu=\lambda$ is the average vehicle arrival rate within the road segment.

$p_e$: probability of transmission failure, that is the probability that one of the vehicles within a coverage area fails in receiving the transmitted information.

$D$: information refresh period, that is the time between two transmissions of the same IR. Information refresh is meaningful for both handling failed transmissions and reaching vehicles that have entered the coverage area during the time D.

A finite state diagram is shown in Fig. 2. The states of the model are indicated by two numbers, denoting the number of users storing the considered IR (i) and the number of users/vehicles which stays within the considered road segment (j), respectively. Clearly, $0 \leq i \leq j \leq N$.

A transition $(i,j) \to (i,j+1)$ represents a vehicle entering the considered road segment. Since the IR is not propagated backward outside the road segment of interest, any vehicle entering the area of interest does not store the considered IR. The relevant transition rate is $\lambda$.

A transition $(i,j) \to (i,j-1)$ represents the exit of a vehicle that does not store the considered IR from the considered road segment. The relevant transition rate is $(j-i) \cdot \mu$.

A transition $(i,j) \to (i-1,j-1)$ represents the exit of a vehicle that stores the considered IR from the considered road segment. The relevant transition rate is $i \cdot \mu$.

A transition $(i,j) \to (i+k,j)$, $i+k \leq j$, represents the delivery of an IR to k further vehicles. In order to find the appropriate transition rate, it is necessary to consider the following aspect. The situation illustrated in Fig.1 is essentially a single dimension problem. As mentioned above, if we assume that $P$ is much higher than the road width, we can approximate the area covered by each transmission as a road segment of length $P$. This implies that every $D$ seconds the information is propagated, on the average, $P/2$ meters forward and $P/2$ meters backward from the vehicle that has initially transmitted the IR. Thus, the average number of vehicles receiving the IR at each time step is the minimum between $N_{ave}$ and j-i. Considering also the probability of transmission failure, and assuming that the failure events are statistically independent, the transition rate can be expressed by resorting to the binomial probability distribution function, as follows:

$$p_{i,j \to i+k,j} = \begin{cases} \frac{1}{D} \binom{N_r}{k} (1-p_e)^k p_e^{N_r-k}, & k \leq N_r \\ 0, & otherwise \end{cases} \quad (1)$$

where $N_r = \min(\lfloor N_{ave} \rfloor, j-i)$ with probability $p_1$ and $N_r = \min(\lceil N_{ave} \rceil, j-i)$ with probability $1-p_1$. We argue that $p_1$ may be evaluated as $\lceil N_{ave} \rceil - N_{ave}$, since $0 \leq \lceil N_{ave} \rceil - N_{ave} < 1$, and the smaller the approximation to the lower integer the higher the probability $p_1$. It should be noted that $N_{ave}$ is an upper limit of the mean value of vehicles that can be reached by the signal transmitted at each step. Therefore, the consequent probability of consistency obtained can be considered an upper limit.

Note also that we have assumed that all vehicles staying within a coverage area can receive the transmitted IR. This is equivalent to say that the probability that a vehicle can exit the coverage area during an IR transmission is negligible. This happens if the size of the IR is very small or when the relative speed of vehicle is so small that the IR transmission time is negligible in comparison with the vehicle sojourn time of the departing state in Fig. 2. It is also obvious that the refresh time must be higher than $D$.

The main approximation of this expression consists of using the memoryless exponential distribution even if the refresh time is deterministic. Nevertheless, given the lack of synchronization between transmitters, we argue that it can be acceptable. In addition, since the standard deviation of the exponential distribution is equal to its mean, when the latter is sufficiently low, the approximation is acceptable.

A further questionable approximation is the memoryless distribution of the vehicle sojourn time within the considered road segment. This approximation is acceptable only if the traffic intensity is not so high to cause queues of vehicles. This approximation works better for short road segments, since the standard deviation of the exponential distribution is equal to its mean, and the actual probability of short vehicle sojourn time converges is well approximated by the assumed model.

For what concerns all other events determining the system evolution, such as a vehicle entering the area of interest, they happen in a substantially random fashion. Hence, their inter-arrival time may be modeled as an exponential random variable.

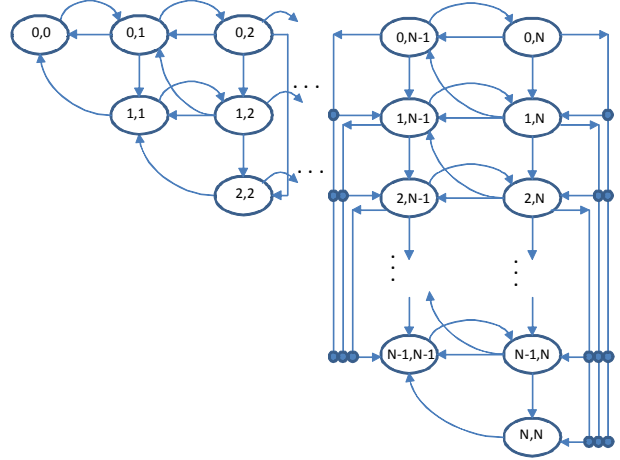

Fig. 2. State diagram of a generic SS IDS

In this paper we evaluate both the transient and the steady state probability distribution of the model states. These probabilities will be used to determine the probability that, at least, a pre-established number $t$ of vehicles can store the desired IR.

All the states shown in Fig. 2 concur to form the *infinitesimal generator*, or *rate matrix*, $\mathbf{A}$ of the state diagram. It is easy to verify that the number of states is $(N^2+N)/2$.

This matrix allows writing the forward *Chapman-Kolmogorov* equations:

$$\frac{d\mathbf{X}(t)}{dt} = \mathbf{X}(t)\mathbf{A} \quad (2)$$

where $\mathbf{X}(t)$ is a $1 \times (N^2+N)/2$ row vector. The solution of (2) is

$$\mathbf{X}(t) = \mathbf{X}_0 e^{\mathbf{A}t} \quad (3)$$

where $\mathbf{X}_0$ is the initial state probability distribution.

The computation of the matrix exponential is prone to severe numerical problems and is still an active research area [5]. In many situation the Laplace transform provides a suitable representation [1]:

$$e^{\mathbf{A}t} = \mathcal{L}^{-1}(s\mathbf{I}-\mathbf{A})^{-1} \quad (4)$$

where $\mathbf{I}$ is the identity matrix and $\mathcal{L}^{-1}$ denotes the inverse Laplace transform. Hence

$$\mathbf{X}(s) = \mathbf{X}_0(s\mathbf{I}-\mathbf{A})^{-1} \quad (5)$$

The steady state probability can be found without inverting the Laplace transform by resorting to the well known *final value theorem*:

$$\lim_{t\to\infty}\mathbf{X}(t)=\lim_{s\to 0}s\mathbf{X}(s) \quad (6)$$

The probability of consistency can be obtained as follows. Given the minimum number $t$ of vehicles to be covered, the states in Fig. 2, in which the system is consistent, are those having the first index $i$ higher or equal to $t$. The number of states in Fig.2 corresponding to less than $t$ covered vehicles is

$$J=\sum_{i=0}^{t-1}(N-i)=tN-\frac{t(t-1)}{2}. \quad (7)$$

Thus, the consistency probability at any given time t can be expressed as:

$$P_{cons}(t)=1-\sum_{i=1}^{J}\mathbf{X}_i(t). \quad (8)$$

It is worth to note that the steady state probability, $P_{cons}=\lim_{t\to\infty}P_{cons}(t)$, can also be expressed as

$$P_{cons}=1-\sum_{i=1}^{J}\mathbf{B}_i \quad (9)$$

Where

$$\mathbf{B}=\frac{null(\mathbf{A}^T)}{\sum_{i=1}^{(N^2+N)/2}null(\mathbf{A}^T)} \quad (10)$$

being $null(\mathbf{A}^T)$ the nullity of the transpose of the rate matrix A. Even if the latter equation does not provide additional results, its usage allows considerable saving of processing load.

### III. CASE STUDIES AND RESULTS

In order to predict the suitability of vehicular systems for data storage, we have analyzed some case studies which differ by some key aspects by using our model.

*Urban Environment*: this situation is characterized by a high vehicle density, travelling at a rather low speed.

*Rural Environment*: this situation is characterized by a moderate vehicle density, travelling at a moderate speed.

*Highway Environment*: this situation is characterized by a low vehicle density, travelling at a high speed.

Table 1 reports the main parameters characterizing the environments defined above, used for the performance evaluation illustrated in what follows. Note that a relatively short communication range allows avoiding many MAC issues. Space limitations do not allow showing further case studies.

**Table 1: parameters describing the considered use cases.**

| Environment | L (m) | V (km/h) | N | $p_e$ | P (m) | D (s) |
|---|---|---|---|---|---|---|
| Urban | 100 | 30 | 10to30 | $10^{-5}$ | 30 | 5 |
| Rural | 500 | 50 | 10to30 | $10^{-5}$ | 30 | 5 |
| Highway | 1000 | 100 | 10to30 | $10^{-5}$ | 30 | 5 |

The performance achievable in the *Urban* environment is shown in Figs 3,4, and 5. Fig. 3 shows the asymptotical probability of storing an IR , in at least $t$ vehicles, versus both $t$ values and number of vehicles located in a 100 m road segment. Fig.4 shows the same probability at 30 seconds from the IR generation. A detailed view of the system's dynamical behavior is shown in Figs. 5(a) to 5(f). Each figure corresponds to a different $t$ value, with $t$=1,5,10,15,20,25, and shows the probability of having an IR stored in at least $t$ vehicles versus a timeframe of 30s since the IR's creation. From Fig. 3 it emerges that, in order to achieve an acceptable probability value, the number of target vehicles $t$ cannot be much higher than half the number of available vehicles. This is clearly an important limitation for safety-related applications.

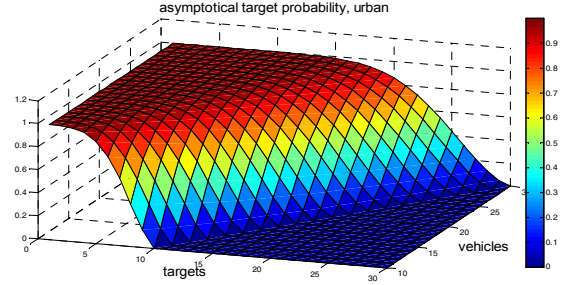

Fig. 3. Steady-state probability of distributing an IR over the desired number of targets, vs. number of targets and vehicles travelling through the road segment of length L. Case study: Urban.

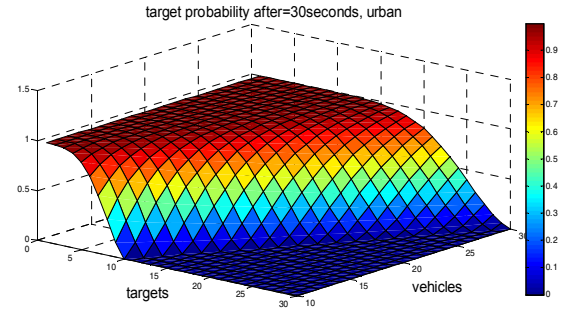

Fig. 4. Probability of distributing an IR over the desired number of targets, after 30 s since an IR arrival, vs. number of targets and vehicles travelling through the road segment of length L. Case study: Urban.

Fortunately, the dynamic behavior is promising. From Fig. 4 it appears that the probability distribution at 30 s from the IR generation is quite close to the limiting one. Figs 5(a) to (f) show that for small $t$ values, 5 to 10 seconds are sufficient to achieve a good storage consistency. Higher $t$ values require a longer time to achieve probability of consistency close to the limiting one. In any case, it emerges that t=20 is a well feasible configuration that allows implementing a wide range of applications. The reason of this result is that the low average speed in the urban environment allows vehicles to stay at a relative short distance. Hence, the vehicle coverage area covers a relatively high number of vehicles and this simplifies information distribution. The performance achievable in the Rural environment is shown in Figs 6,7, and 8. The organization of the displayed information is similar the one on the Urban environment. From Fig. 6 it emerges that, regarding the asymptotical probability, results achieved in the rural environment are similar to those achieved in the urban one. This behavior is quite intuitive since the state-state probability is highly dependent on the number of available vehicles. Nevertheless, from the subsequent figures a substantially different dynamical behavior emerges. For most of analyzed situations, 30 seconds are not sufficient to get close enough to the asymptotical probability values. In addition, also for small t values, the curves are less convergent for any number of considered vehicles. This behavior is essentially due to the increased vehicle speed, which increases the inter-vehicle distance due to safety reasons.

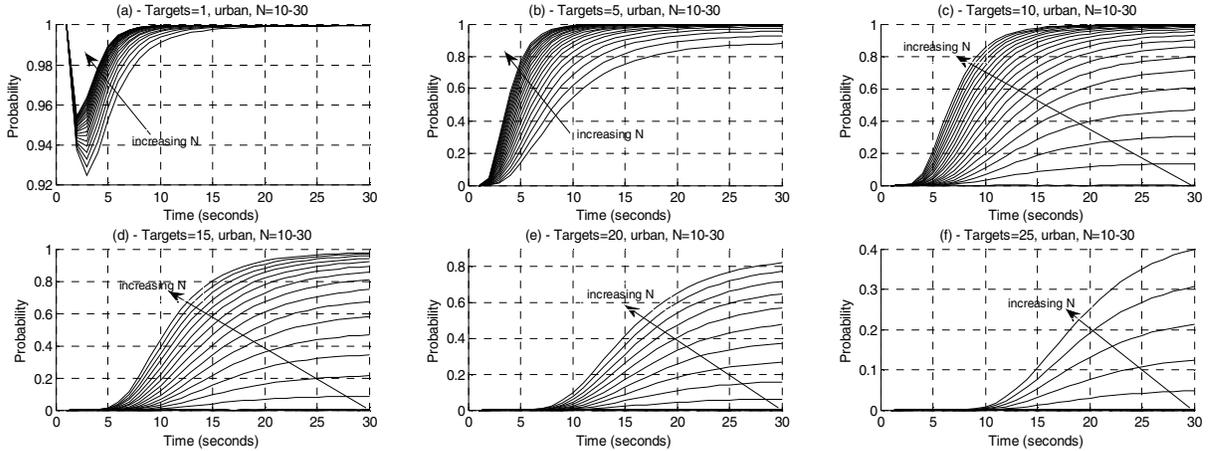

Fig. 5. Probability of having an IR stored in t=1 (a), 5(b), 10(c), 15(d), 20(e), 25(f) vehicles vs. time since the IR arrival time. Case study: Urban.

Since the coverage area is the same as in the previous case study, it emerges that the average number of vehicles staying within a coverage area decreases, as well as the information distribution speed.

The performance achievable in the Highway environment is shown in Figs 9,10, and 11, with the same organization of contents of the previously examined case studies. From Fig. 9 it emerges again that the asymptotical probability is highly dependent on the number of available vehicles, and the probability distribution is very close to the one of the previous case study. Nevertheless, the increased average speed of vehicles makes the dynamic behavior quite critical, also for small $t$ values. Fig. 10 shows that at 30 seconds from the IR generation, the plotted surface is flattened for t>10. Even for $t$=5, 30 seconds are not sufficient to have $t$ vehicles covered with a good probability if the number of available vehicles is lower than 25. Furthermore, Figs 11 (b) to (f) shows that tens of seconds are necessary even just to move the IR between vehicles. Consider that the maximum number of vehicles considered is 30 within a road length of 1km. Thus, the traffic conditions cannot be regarded as "very light traffic". This implies that the Highway environment is extremely critical for the introduction of services which need promptness in information diffusion.

## IV. CONCLUSIONS

In this paper we have proposed a Markovian model for analyzing the suitability of using vehicular devices for establishing data storage systems.

For this purpose, we have analyzed three different environments, namely a urban environment, a rural environment, and a highway environment. We have shown the convergence properties of the proposed model to the steady state distribution analytically. The vehicle traffic conditions that we have considered are such to include a significant number of vehicles, but not so high to cause congestion and queues.

Our analysis shows that by using a communication range of some tens of meters only the urban environment, being characterized by the highest vehicle density, shows a dynamic behavior compliant with applications requesting high responsiveness to information updates. On the contrary, the high latency of the highway environment limits its suitability to the distribution of stationary information types.

Future research directions include a detailed protocol architecture and real experiments on a vehicular network.

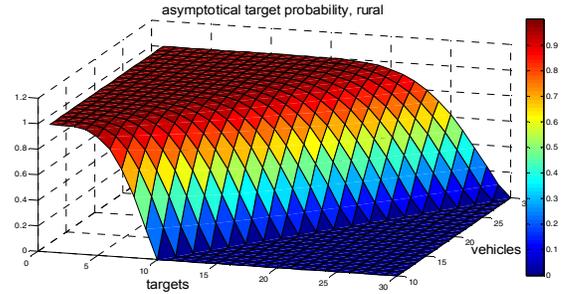

Fig. 6. Steady-state probability of distributing an IR over the desired number of targets, vs. number of targets and vehicles travelling through the road segment of length L. Case study: Rural.

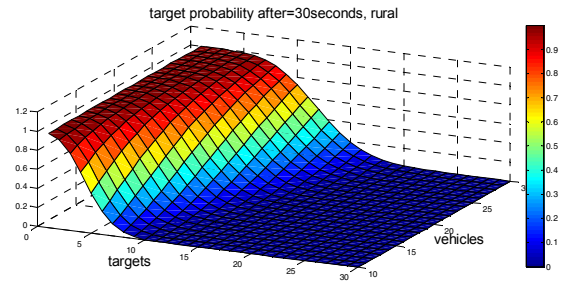

Fig. 7. Probability of distributing an IR over the desired number of targets, after 30 s since an IR arrival, vs. number of targets and vehicles travelling through the road segment of length L. Case study: Rural.

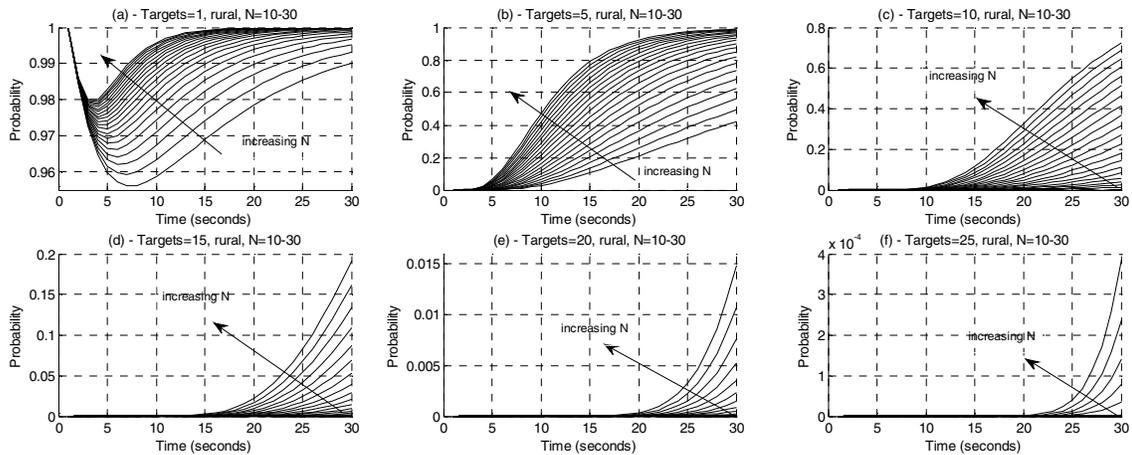

Fig. 8. Probability of having an IR stored in t=1 (a), 5(b), 10(c), 15(d), 20(e), 25(f) vehicles vs. time since the IR arrival time. Case study: Rural.

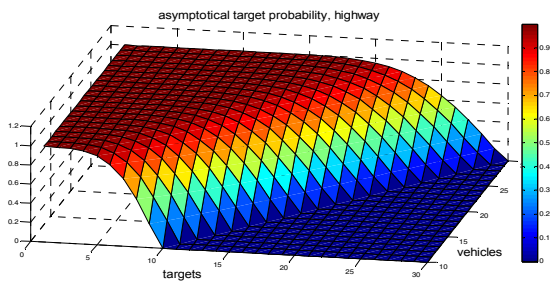

Fig. 9. Steady-state probability of distributing an IR over the desired number of targets, vs. number of targets and vehicles travelling through the road segment of length L. Case study: Highway.

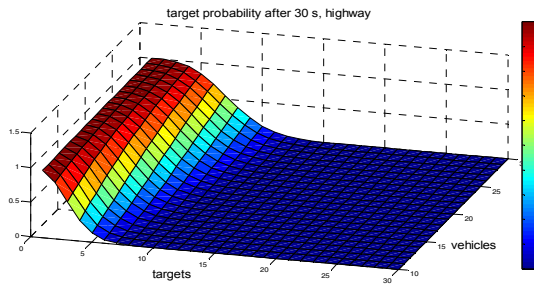

Fig. 10. Probability of distributing an IR over the desired number of targets, after 30 s since an IR arrival, vs. number of targets and vehicles travelling through the road segment of length L. Case study: Highway.

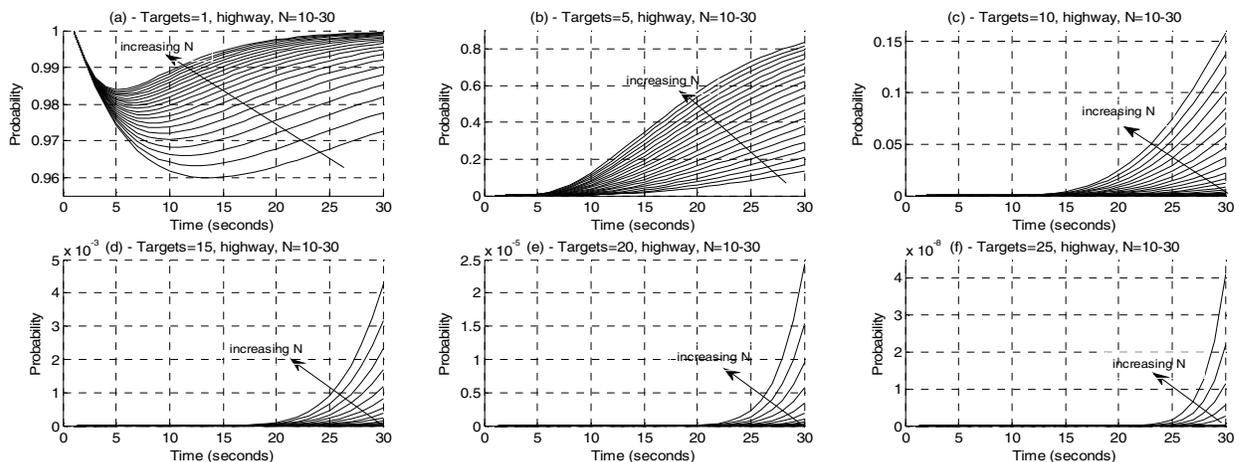

Fig. 11. Probability of having an IR stored in t=1 (a), 5(b), 10(c), 15(d), 20(e), 25(f) vehicles vs. time since the IR arrival time. Case study: Highway.